\documentstyle[epsf]{article}
\begin{document}
\onecolumn
\begin{titlepage}
\begin{center}
{\Large \bf Exact Solution for Relativistic Two-Body Motion\\
in Dilaton Gravity} \\ \vspace{2cm}
R.B. Mann\footnotemark\footnotetext{email: 
mann@avatar.uwaterloo.ca}\\
\vspace{1cm}
Dept. of Physics,
University of Waterloo
Waterloo, ONT N2L 3G1, Canada\\
and \\
\vspace{1cm}
T. Ohta \footnotemark\footnotetext{email:
t-oo1@ipc.miyakyo-u.ac.jp}\\
\vspace{1cm} 
Department of Physics, Miyagi University of Education,
Aoba-Aramaki, Sendai 980, Japan\\
\vspace{2cm}
\vspace{2cm}
\today\\
\end{center}
\begin{abstract}

We present an exact solution to the problem
of the relativistic motion of
2 point masses in $(1+1)$ dimensional dilaton gravity. 
The motion of the bodies is governed entirely by their
mutual gravitational influence, and the spacetime metric is likewise
fully determined by their stress-energy. A Newtonian limit exists,
and there is a static gravitational potential.  
Our solution gives the exact Hamiltonian to infinite order in the 
gravitational coupling constant. 

\end{abstract}
\end{titlepage}
\onecolumn

The problem of motion is a notoriously difficult one in gravitational
theory. Although approximation techniques exist  \cite{300yrs},
in general there is no exact solution to the
problem of the motion of $N$ bodies each interacting under their
mutual gravitational influence, except in the case $N=2$ for 
Newtonian gravity, or in $(2+1)$ dimensions, where
the absence of a static gravitational potential 
allows one to generalize the static 2-body metric to that of two
bodies moving with any speed \cite{Bellini}.  

We present here an exact solution to problem of the relativistic motion of
2 point masses under gravity in $(1+1)$ dimensions. 
 The dimensionality
necessitates that the gravitational theory we choose is a dilaton theory
of gravity; however our choice of dilatonic gravity is such that the 
dilaton decouples from the classical equations of motion \cite{RT}.  
Consequently the motion of the bodies is governed entirely by their
mutual gravitational influence, and the spacetime metric is likewise
fully determined by their stress-energy \cite{RT2}. 
Unlike the $(2+1)$ dimensional case, a Newtonian limit exists,
and there is a static gravitational potential.  
Our solution gives the exact Hamiltonian to infinite order in the 
gravitational coupling constant. We can thus view the whole structure of 
the theory from the weak field to the strong field limits.

We begin with the $N$-body gravitational action
\begin{eqnarray}\label{act1}
I&=&\int d^{2}x\left[
\frac{1}{2\kappa}\sqrt{-g}
\left\{\Psi R+\frac{1}{2}g^{\mu\nu}\nabla_{\mu}\Psi\nabla_{\nu}\Psi\right\}
\right.
\nonumber \\
&&\makebox[2em]{}-\left.\sum_{a=1}^N m_{a}\int d\tau_{a}
\left\{-g_{\mu\nu}(x)\frac{dz^{\mu}_{a}}{d\tau_{a}}
\frac{dz^{\nu}_{a}}{d\tau_{a}}\right\}^{1/2}\delta^{2}(x-z_{a}(\tau_{a}))
\right] 
\end{eqnarray}
\noindent
where $\Psi$ is the  dilaton field and $g_{\mu\nu}$,
$g$, are the metric and its determinant, $R$ is the Ricci
scalar and $\tau_{a}$  is the proper time of $a$-th particle, 
with $\kappa=8\pi G/c^4$. It is straightforward to show that the
system of field equations reduces to
\begin{equation}\label{RTgeo}
R=\kappa T^{\mu}_{\;\;\mu} \qquad
\frac{d}{d\tau_{a}}
\left\{\frac{dz^{\nu}_{a}}{d\tau_{a}}\right\}
-\Gamma^\nu_{\alpha\beta}(z_a)
\frac{dz^{\alpha}_{a}}{d\tau_{a}}
\frac{dz^{\beta}_{a}}{d\tau_{a}}=0 \;\;.
\end{equation}
and
\begin{equation}\label{e4}
\frac{1}{2}\nabla_{\mu}\Psi\nabla_{\nu}\Psi
-g_{\mu\nu}\left(\frac{1}{4}\nabla^{\lambda}\Psi\nabla_{\lambda}\Psi
-\nabla^{2}\Psi\right)
-\nabla_{\mu}\nabla_{\nu}\Psi=\kappa T_{\mu\nu} 
\end{equation}
where the stress-energy due to the point masses is
\begin{equation}\label{stress}
T_{\mu\nu} = \sum_{a=1}^N m_{a}\int d\tau_{a}\frac{1}{\sqrt{-g}}
g_{\mu\sigma}g_{\nu\rho}\frac{dz^{\sigma}_{a}}{d\tau_{a}}
\frac{dz^{\rho}_{a}}{d\tau_{a}}\delta^{2}(x-z_{a}(\tau_{a}))\;\;  ,
\end{equation}
and is conserved.  Note that (\ref{RTgeo}) is a closed system of 
$N+1$ equations for which one can solve for the single metric degree
of freedom and the $N$ degrees of freedom of the point masses. The
evolution of the dilaton field is governed by the evolution of the
point-masses via (\ref{e4}). It is easy to show that the left-hand side
of this equation is divergenceless (consistent with the conservation
of $T_{\mu\nu}$), yielding only one independent equation to determine
the single degree of freedom of the dilaton.

We shall work in the canonical formalism for which the action
(\ref{act1}) is written in the form\cite{ohtarobb}
\begin{equation}\label{e9}
I=\int dx^{2}\left\{\sum_{a}p_{a}\dot{z}_{a}\delta(x-z_{a}(x^{0}))
+\pi\dot{\gamma}+\Pi\dot{\Psi}+N_{0}R^{0}+N_{1}R^{1}\right\} 
\end{equation}
where  $\gamma=g_{11},  N_{0}= (-g^{00})^{-1/2}, N_{1}= g_{10}$,
$\pi$ and $\Pi$ are conjugate momenta to $\gamma$ and $\Psi$
respectively, and 
\begin{eqnarray}
R^{0}&=&-\kappa\sqrt{\gamma}\gamma\pi^{2}+2\kappa\sqrt{\gamma}\pi\Pi
+\frac{1}{4\kappa\sqrt{\gamma}}(\Psi^{\prime})^{2}
-\frac{1}{\kappa}\left(\frac{\Psi^{\prime}}{\sqrt{\gamma}}\right)^{\prime}
-\sum_{a}\sqrt{\frac{p^{2}_{a}}{\gamma}+m^{2}_{a}}\;
\delta(x-z_{a}(x^{0}))
\nonumber \\
\\
R^{1}&=&\frac{\gamma^{\prime}}{\gamma}\pi-\frac{1}{\gamma}\Pi\Psi^{\prime}
+2\pi^{\prime}
+\sum_{a}\frac{p_{a}}{\gamma}\delta(x-z_{a}(x^{0})) \;\;.
\end{eqnarray}
with the symbols $(\;\dot{}\;)$ and  $(\;^{\prime}\;)$
denoting $\partial_{0}$ and $\partial_{1}$, respectively.

The quantities $N_0$ and $N_1$ are Lagrange multipliers which enforce
the constraints $R^0 = 0 = R^1$.  Since the only linear terms 
in these constraints are $(\Psi^{\prime}/\sqrt{\gamma})^{\prime}$ 
and $\pi^{\prime}$, we may solve for these quantities in terms of the 
dynamical and gauge ({\it i.e.} co-ordinate) degrees of freedom. 
These latter degrees of freedom may be identified by writing the generator 
(which arises from the variation of the action at the boundaries)
in terms of the former quantities, and then finding which
quantities serve to fix the frame of the physical space-time 
coordinates in a manner analogous
to the $(3+1)$-dimensional case \cite{r7,r8,adm}. 

Carrying out this procedure,
we find that we can consistently choose the coordinate conditions
$\gamma=1$ and $\Pi=0$. Eliminating the constraints, the action 
(\ref{e9}) then reduces to 
\begin{equation}
I=\int d^{2}x\left\{\sum_{a}p_{a}\dot{z}_{a}\delta(x-z_{a})
-\cal H\mit\right\}\;\;.
\end{equation}
where the reduced Hamiltonian for the system of particles is 
$H=\int dx \cal H\mit =-\frac{1}{\kappa}\int dx \triangle\Psi$ ,
where $\triangle \equiv \partial^{2}/\partial x^{2}$.

Here $\Psi = \Psi(x,z_{a},p_{a})$ and is determined
by solving the constraint equations which are now
\begin{equation}\label{psicon}
\triangle\Psi-\frac{1}{4}(\Psi^{\prime})^{2}
+\kappa^{2}\pi^{2}+\kappa\sum_{a}\sqrt{p^{2}_{a}+m^{2}_{a}}
\delta(x-z_{a})=0 \;\;,
\end{equation}
\begin{equation}\label{picon}
2\pi^{\prime}+\sum_{a}p_{a}\delta(x-z_{a})=0 \;\;.
\end{equation}

An exact solution to these equations in the 2-body case
may now be obtained as follows.
Consider first the case $z_{2}<z_{1}$, for which we may divide 
spacetime into three regions: $z_{1}<x$ ((+) region), 
$z_{2}<x<z_{1}$ ((0) region) and $x<z_{2}$ ((-) region). Writing
$\Psi=-4\mbox{log}|\phi|$ and $\pi=\chi^\prime$
we find in each region that 
$\phi$ is the sum of 
growing and decaying exponentials in $x$ and that
$\chi$ is a sum of terms linear in $x$. Matching these
solutions at the boundaries $x=z_1$ and $x=z_2$ of each region
allows a determination of the coefficients of the exponentials
in the $+$ and $-$ regions in terms of those in the $0$ region.

Because the magnitudes of both $\phi$ and $\chi$ increase with 
increasing $|x|$, it is necessary to impose a boundary condition
which guarantees that the surface terms obtained in passing from
(\ref{act1}) to (\ref{e9}) vanish. This condition has been shown
to be $\Psi^{2}-4\kappa^{2}\chi^{2}=0$, which must hold in 
the $+$ and $-$ regions \cite{ohtarobb}.  Incorporating this into the matching
conditions allows one to fully determine the coefficients of
the exponentials in the $0$ region in terms of the momenta and
positions of the point masses.  

Solving the constraints and the equations for $N_{0}$ and $N_{1}$
yields
\begin{equation}
N_{0}=Ae^{-\frac{1}{2}\Psi}=A\phi^{2}=\left\{
\begin{array}{ll}
A\phi^{2}_{+} & \makebox[3em]{}(+)\mbox{region} \\
A\phi^{2}_{0} & \makebox[3em]{}(0)\mbox{region} \\
A\phi^{2}_{-} & \makebox[3em]{}(-)\mbox{region}
\end{array}
\right.
\end{equation}
and
\begin{eqnarray}
N_{1(+)}=\epsilon\left(A\phi^{2}_{+}-1\right)
&\qquad&
N_{1(-)}=-\epsilon\left(A\phi^{2}_{-}-1\right)
\end{eqnarray}
\begin{equation}
N_{1(0)}=\epsilon A\frac{L_{1}L_{2}}{L^{2}_{0}}\left\{
\frac{L_{2}}{M_{1}}e^{\frac{\kappa}{4}L_{0}(x-z_{1})}
-\frac{L_{1}}{M_{2}}e^{-\frac{\kappa}{4}L_{0}(x-z_{2})}\right\}
+\frac{\kappa\epsilon}{2}A\frac{L_{1}L_{2}}{L_{0}}x+\epsilon C_{0}
\end{equation}
where $A={L_{0}}/{(L_{1}+L_{2}-\frac{\kappa}{4}(z_1-z_2)L_{1}L_{2})}$,
$C_{0}=A(M_{1}-M_{2}-\frac{\kappa}{4}(z_{1}+z_{2})L_{1}L_{2})/L_0$,
$\epsilon^2=1$, and
\begin{eqnarray}
\phi_{+}=\left(\frac{L_{1}}{M_{1}}\right)^{1/2}e^{\frac{\kappa}{8}
L_{+}(x-z_{1})}
&\qquad&
\phi_{-}=\left(\frac{L_{2}}{M_{2}}\right)^{1/2}e^{-\frac{\kappa}{8}
L_{-}(x-z_{2})}
\end{eqnarray}
\begin{equation}
\phi_{0}=\frac{(L_{1}L_{2})^{1/2}}{L_{0}}\left[
\left(\frac{L_{2}}{M_{1}}\right)^{1/2}e^{\frac{\kappa}{8}L_{0}(x-z_{1})}
+\left(\frac{L_{1}}{M_{2}}\right)^{1/2}e^{-\frac{\kappa}{8}L_{0}(x-z_{2})}
\right]
\end{equation}
as the solutions for the relevant field variables.
For convenience we have defined the quantities
\begin{eqnarray}
M_{a}\equiv \sqrt{p^{2}_{a}+m^{2}_{a}}+\epsilon\eta_{ab} p^{b}
&\quad& L_{\pm} = 4X \pm\epsilon(p_1+p_2)\\
L_{1}\equiv L_0 - M_{2} &\quad&  L_{2}\equiv L_0 - M_{1} 
\end{eqnarray}
where $-\eta_{11}=\eta_{22}=1, \eta_{12}=\eta_{21}=0$
and $L_{0} \equiv  4X-\epsilon (p_{1}-p_{2})$. These 
quantities are related by the equation
\begin{equation}\label{eqX}
L_{1}L_{2}=M_{1}M_{2}e^{\frac{\kappa}{4}L_{0}(z_{1}-z_{2})}
\end{equation}
which determines $X$.

The parameter $\epsilon$ is a constant of integration associated
with the  metric degree of freedom. We have two types of solutions 
corresponding to $\epsilon=\pm 1$. Under time reversal, these
solutions transform into each other, ensuring invariance of the whole 
theory under this symmetry. 
It is straightforward to show from this solution
that the Ricci scalar vanishes everywhere except at the locations 
$(z_1(t),z_2(t))$ of the point masses.

A calcuation shows that the Hamiltonian  of the system is
$H=4X$. Consequently (\ref{eqX}) gives an exact solution
for the Hamiltonian $H$ in terms of the co-ordinate
and metric degrees of freedom. It is straightforward to show that the 
total momenta $p_1 + p_2$ is conserved, and so we may choose a frame of 
reference with $p_1=-p_2=p$. Hamilton's equations then yield
\begin{eqnarray}
\dot{p}&=&-\frac{\kappa}{4} A L_{1}L_{2}\label{eqnp}\\
\dot{z_{1}}&=&\epsilon- \epsilon A \frac{L_{1}}{\sqrt{p^{2}+m^{2}_{1}}}
\label{eqnz1}\\
\dot{z_{2}}&=&-\epsilon +\epsilon A \frac{L_{2}}{\sqrt{p^{2}+m^{2}_{2}}}
\label{eqnz2}
\end{eqnarray}
as the dynamical equations for the 2-body system coupled to gravity.

Repeating this analysis for $z_{1}<z_{2}$ yields the general equation
for the Hamiltonian
\begin{eqnarray}
&&(\sqrt{p^{2}_{1}+m^{2}_{1}}-\epsilon\tilde{p}_{2}-H)
(\sqrt{p^{2}_{2}+m^{2}_{2}}+\epsilon\tilde{p}_{1}-H)
\nonumber \\
&& \makebox[5em]{}=(\sqrt{p^{2}_{1}+m^{2}_{1}}-\epsilon\tilde{p}_{1})
(\sqrt{p^{2}_{2}+m^{2}_{2}}+\epsilon\tilde{p}_{2})
e^{\frac{\kappa}{4}\left\{H-\epsilon(\tilde{p}_{1}-\tilde{p}_{2})\right\}\;|r|}
\label{ham}
\end{eqnarray}
where $r\equiv z_{1}-z_{2}$ and 
$\tilde{p}_{a} \equiv p_{a} \mbox{sgn}{(z_1-z_2)}$.

Equation (\ref{ham}) describes the surface in $(r,p,H)$ space of all
allowed phase-space trajectories. Since  $H$ is a constant of the motion, 
(a fact easily verified by differentiation of (\ref{ham}) 
with respect to $t$) a given trajectory in the $(r,p)$ plane
is uniquely determined by setting  $H=H_0$ in (\ref{ham}).  
For $H_0$ sufficiently small, the trajectory
can be regarded as a relativistic perturbation of the Newtonian case.
However once $H_0$ is sufficiently large there exist a qualitatively 
new set of trajectories which cannot be understood as relativistically
`correcting' the Newtonian ones. The transition from the first set 
to the second set is smooth.

In the equal-mass case, the Hamiltonian is
\begin{equation}\label{eqmass}
H =  - 8\,\frac{W\left(-\frac{\kappa}{8}
 \left( \! |r|\,\sqrt{{p}^{2}+ m^2} - {\epsilon{p} r}\, \!  \right) 
\exp\left[{\frac {\kappa}{8}}
 \left( \! |r|\,\sqrt{{p}^{2}+ m^2} - 
\epsilon{p}\, r \right)\right]\right)}{\kappa|r|} + \sqrt{{p}^{2}+ m^2}
+ \epsilon p\, \mbox{sgn}(r)
\end{equation}
where $W(x)$ is the Lambert W-function defined via
\begin{equation}\label{lamb}
y\cdot e^{y}=x \qquad \Longrightarrow \qquad y=W(x)
\end{equation}
and has two real branches \cite{lambert} which join
smoothly onto each other. The principal branch
(for which $W(x) \ge -1$) in (\ref{eqmass}) reduces to the Newtonian
limit for small $\kappa$; the second set of trajectories mentioned
above are described by the other branch.

Some characteristic phase-space plots for the equal mass case
are given in figs. 1--2. In each of
these we have included the corresponding trajectory from the Newtonian
theory for comparison.  We see that as $H_0$ increases, the trajectory
becomes more `S'-shaped, with the particles reaching their maximum 
separation for some positive value of $p$, where the velocity $\dot{r}=0$.  
This occurs because $p$ depends upon the momenta of the particles and of 
the metric.  Under time-reversal, the trajectory for a given value 
of $H_0$ is obtained by reflection in the $r=0$ axis.  Fig. 3 shows 
three phase space plots for the unequal mass case, with the mass of particle
2 being $1/10$, equal to and twice the mass of particle 1.

An interesting limiting case of (22) may be obtained by setting 
$z_{1}=z, z_{2}=0, m_{1}=\mu, m_{2}=m, p_{1}=p$ and $p_{2}=0$. Taking the 
lowest order terms of $\mu$, we obtain
\begin{equation}
H=m+\sqrt{p^{2}+\mu^{2}}e^{\frac{\kappa m}{4}|z|}
-\epsilon p\; \mbox{sgn}(z)\left(e^{\frac{\kappa m}{4}|z|}-1\right)\;\;.
\end{equation}
which is just the Hamiltonian of a test particle of mass $\mu$ under the 
influence of a static source at the origin of mass $m$, 
in which the metric is $N_{0}=e^{\frac{\kappa m}{4}|z|}$, $N_{1}=\epsilon\; \mbox{sgn}(z)\left(
e^{\frac{\kappa m}{4}|z|}-1\right)$ \cite{ohtarobb}. 

We hope to extend our analysis of this problem to the
many-body problem and to more general theories of dilaton gravity.

\section*{Acknowledgements} This work was supported in part by the
Natural Sciences and Engineering Research Council of Canada.

\vspace{2cm}

\section*{Figure Captions}

\noindent
{\bf Fig. 1}\hspace{.5 cm} A comparison of relativistic (s-shaped) and 
non-relativistic (oval) phase-space trajectories for $H=2.2m$. All
axes are in units of $m$.

\noindent
{\bf Fig. 2}\hspace{.5 cm} A comparison of relativistic (s-shaped) and 
non-relativistic (oval) phase-space trajectories for $H=4m$. All
axes are in units of $m$.

\noindent
{\bf Fig. 3}\hspace{.5 cm} Phase-space trajectories for $H=4m_1$, with
$m_2 = 2m_1$ (innermost curve), $m_2=m_1$ and $m_2=.1m_1$ (outermost
curve). All axes are in units of $m_1$.

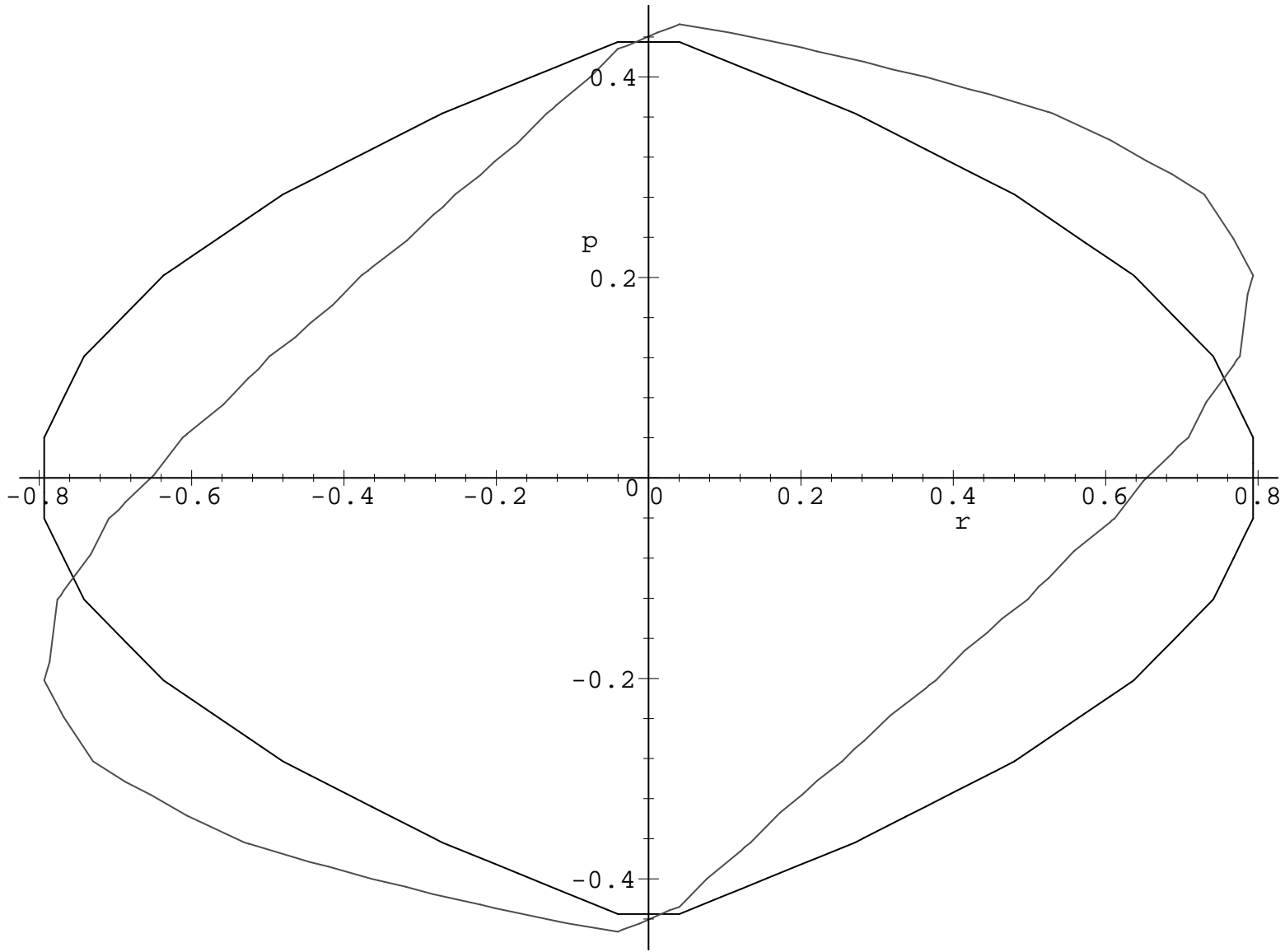
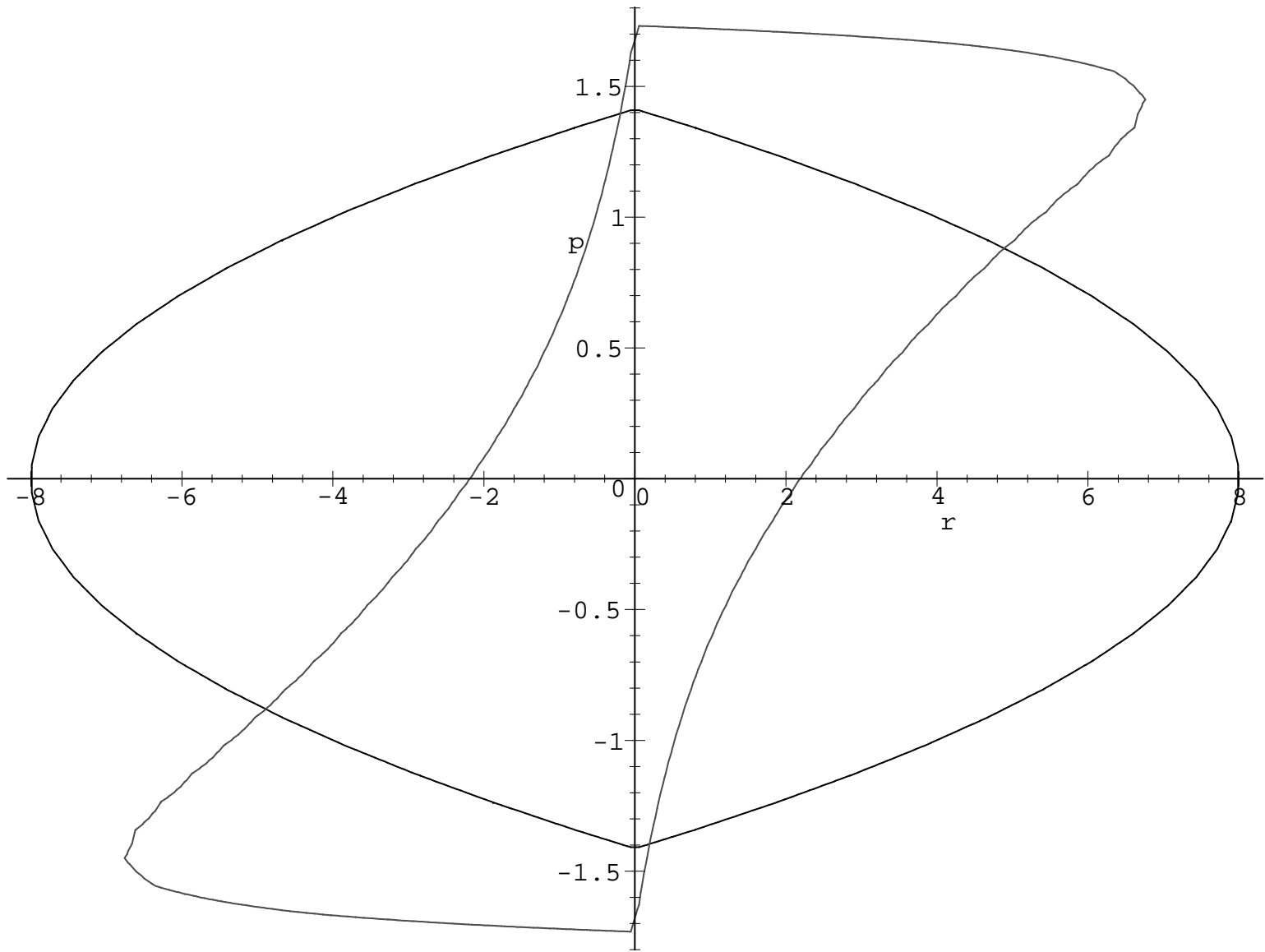
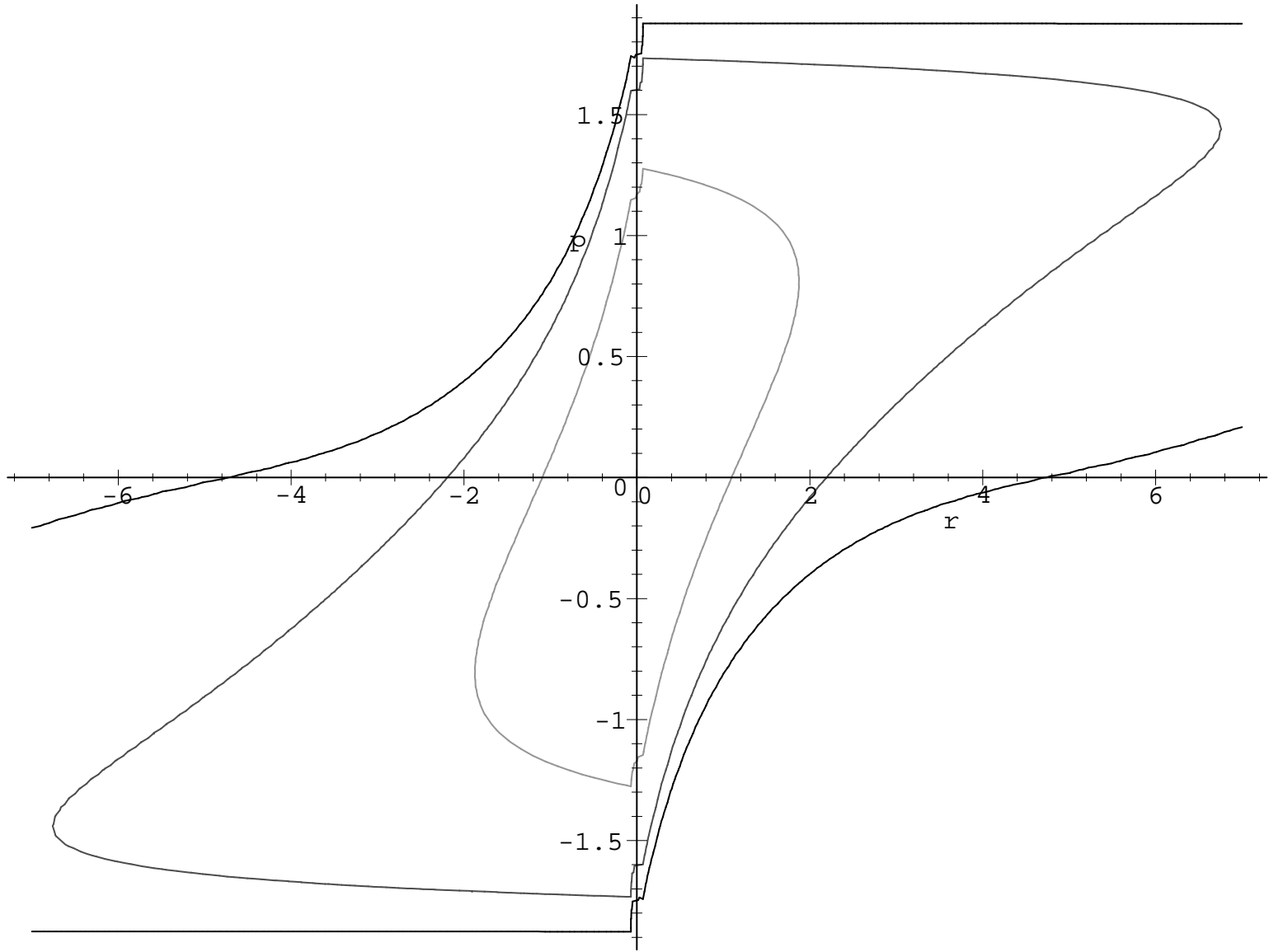
\end{document}